\documentclass[a4paper,11pt]{article}
\pdfoutput=1
\usepackage{jheppub}

\usepackage[utf8]{inputenc}
\usepackage{graphicx,rotating,amsmath,amsfonts,amssymb}
\usepackage{enumerate}

\usepackage{tikzexternal}
\usepackage{tikz}
\usetikzlibrary{external}

\newcommand{\be}{\begin{equation}}
\newcommand{\ee}{\end{equation}}
\newcommand{\bea}{\begin{equation}\begin{aligned}} 
\newcommand{\eea}{\end{aligned}\end{equation}}

\newcommand{\GeV}{{\rm GeV}}

\newcommand{\mpl}{M_P}

\newcommand{\sv}{\langle\sigma v\rangle}
\newcommand{\Trh}{T_\text{RH}}
\newcommand{\Tmax}{T_\text{max}}
\newcommand{\gs}{g_\star}
\newcommand{\gss}{g_{\star s}}
\newcommand{\gskin}{g_\star^{\rm kin}}
\newcommand{\gsrh}{g_\star^{\rm RH}}
\newcommand{\gsskin}{g_{\star s}^{\rm kin}}
\newcommand{\gssrh}{g_{\star s}^{\rm RH}}

\newcommand{\Hkin}{H_{\rm kin}}

\title{Boosting Ultraviolet Freeze-in in NO Models}

\author[a]{Nicol\'as Bernal,}
\author[b]{Javier Rubio}
\author[c]{and Hardi Veerm\"ae}

\affiliation[a]{Centro de Investigaciones, Universidad Antonio Nariño\\
Carrera 3 Este \# 47A-15, Bogotá, Colombia}
\affiliation[b]{Department of Physics and Helsinki Institute of Physics\\
PL 64, FI-00014 University of Helsinki, Finland}
\affiliation[c]{National Institute of Chemical Physics and Biophysics\\
R\"avala 10, 10143 Tallinn, Estonia}

\emailAdd{nicolas.bernal@uan.edu.co}
\emailAdd{javier.rubio@helsinki.fi}
\emailAdd{hardi.veermae@cern.ch}

\abstract{ 
We present a novel UV freeze-in mechanism for enhancing the dark matter abundance in cosmologies where inflation is followed by an epoch dominated by a fluid stiffer than radiation. In such scenarios, even a small radiation abundance, produced for instance by instantaneous preheating effects, will eventually dominate the total energy density of the Universe without the need for a complete inflaton decay. For the sake of concreteness, we focus in non-oscillatory (NO) quintessential inflation models, albeit our treatment is rather general and can be extended to other scenarios. The high temperature of the initial thermal bath together with the absence of subsequent entropy injections into the Standard Model plasma translates into a highly-effective UV DM freeze-in. In particular, we find that, during kination, an enhancement in the DM abundance is generically obtained for any production cross section not-decreasing with temperature.}

\begin{document}
\begin{flushright}
    PI/UAN-2020-670FT \\ 
    HIP-2020-12/TH
\end{flushright}

\maketitle 

\section{Introduction}
\label{sec:in}

The existence of a dark matter (DM) component has been firmly established by astrophysical and cosmological observations, although its fundamental nature remains elusive~\cite{Bertone:2016nfn}. Extensive experimental searches for non-gravitational interactions between the dark sector and the Standard Model (SM) have so far produced null results. This has motivated the quest of particle DM models beyond the standard paradigm of Weakly Interacting Massive Particles (WIMP)~\cite{Jungman:1995df, Arcadi:2017kky}.

In the standard WIMP scenario, DM is a thermal relic produced by the freeze-out mechanism. However, the observed DM abundance may be generated also out of equilibrium by the so-called freeze-in mechanism~\cite{McDonald:2001vt, Choi:2005vq, Kusenko:2006rh, Petraki:2007gq, Hall:2009bx} (for a recent review, see Ref.~\cite{Bernal:2017kxu}). Here we focus on a particular subcase, the so-called ultraviolet (UV) freeze-in scenario \cite{Elahi:2014fsa}, where the temperature of the thermal bath is always lower than the mass of the mediator states connecting the DM species to the SM sector.

DM generation from UV freeze-in requires the cross section for DM production via interactions with the SM thermal plasma to grow with increasing temperature. A theoretically well--motivated parametrization of the thermally averaged cross section is~\footnote{For an extensive list of scenarios giving rise to such cross sections see e.g. Ref.~\cite{Bernal:2019mhf}.}
\be\label{eq:sv}
	\sv=\frac{T^n}{\Lambda^{n+2}}\,,
\ee
with $T$ the photon temperature and $\Lambda$ a dimensionful quantity encoding the scale of new physics. This cross section generates a relic abundance of the form $Y\propto \mpl\,\Trh^{n+1}/\Lambda^{n+2}$, with $\Trh$ the so-called reheating temperature at the onset of radiation domination, which, assuming an instantaneous and complete inflaton decay, coincides with the highest temperature of the thermal bath. The strong $\Trh$ dependence in this expression characterizes the UV freeze-in production.

In a more realistic picture in which the instantaneous decay approximation is not used for reheating, the SM bath temperature may initially display a temperature $\Tmax$ much higher than 
$\Trh$ at onset of the radiation domination~\cite{Giudice:2000ex}.
If the processes connecting the visible and DM sectors grows sufficiently fast with temperature, then the bulk of the DM relic abundance might be generated during the heating period, i.e. for $\Tmax > T > \Trh$.
The DM relic abundance would be then established during a stage in which the Universe is not dominated by the SM thermal bath (see e.g.~\cite{McDonald:1989jd, Chung:1998rq, Giudice:2000ex, Allahverdi:2002nb, Allahverdi:2002pu, Gelmini:2006pw, Garcia:2017tuj}). 
For example, if the DM component is produced during the transition from matter to radiation domination via an effective operator with a temperature dependence $n>6$ in Eq.~\eqref{eq:sv}, the DM abundance is enhanced by a ``boost factor'' $B\propto(\Tmax/\Trh)^{n-6}$ as compared to UV freeze-in assuming instantaneous decay~\cite{Garcia:2017tuj}. Subsequent works have explored the impact of this boost factor in specific models~\cite{Chen:2017kvz, Bernal:2018qlk, Bhattacharyya:2018evo, Chowdhury:2018tzw, Kaneta:2019zgw, Banerjee:2019asa, Dutra:2019xet, Dutra:2019nhh, Mahanta:2019sfo, Bernal:2019mhf, Cosme:2020mck, Garcia:2020eof, Bernal:2020fvw, Bernal:2020qyu}.

Recently, it has been observed that the boost factor $B$ depends on the details of cosmological expansion during the heating stage and thus on the shape of the inflaton potential~\cite{Bernal:2019mhf, Garcia:2020eof}.
In detail, when the exponent $n$ in Eq.~\eqref{eq:sv} exceeds a critical value $n_c$, the boost factor takes the form
\begin{equation}\label{eq:bf}
	B \propto \left( \frac{\Tmax}{\Trh} \right)^{n-n_c}.
\end{equation}
If heating takes place through the decay of the inflaton field during a cosmological epoch characterized by the equation-of-state parameter $w$, then $n_c = 2\,(3-w)/(1+w)$. Thus, quadratic and quartic inflaton potentials correspond to $n_c=6$ and $n_c=4$, respectively.

In this paper we present a novel mechanism for enhancing the DM abundance in UV freeze-in scenarios.
We show that if the Universe becomes dominated after the end of inflation by a fluid component with an equation-of-state stiffer than that of radiation, i.e. $w>1/3$, then the DM relic abundance is boosted by a factor displaying the same parametric dependence as the one in Eq.~\eqref{eq:bf} but a lower critical exponent
\be
    n_c = \frac32(w-1)\,.
\ee
Moreover, since the background fluid is diluted faster than radiation, it will naturally become subdominant at temperatures $T<\Trh$, without the need of considering a full depletion of its energy density into SM radiation. This avoids additional entropy injections into the SM particles and therefore the dilution of the produced DM population, maximizing with it the boost factor.
In fact, for $1/3<w\leq 1$, the DM abundance generated by {\it any} higher-order operator (with dimension equal or bigger than 5) gets parametrically enhanced by a factor $\propto (\Tmax/\Trh)^{n-n_c}$.

For the sake of concreteness, this paper focuses on non-oscillatory (NO) quintessential inflation models~\cite{Spokoiny:1993kt, Peebles:1998qn, Brax:2005uf, Hossain:2014xha, Dimopoulos:2017zvq,Rubio:2017gty, Geng:2017mic, Agarwal:2017wxo,Akrami:2017cir} for which the inflation potential does not possess a minimum and inevitably enters a kination dominated regime soon after the end of inflation.%
\footnote{Reference~\cite{Garcia:2020eof} presents a related analysis, focusing on power-law inflationary potentials rather than on NO scenarios. As we will see, this leads to important differences between the two approaches.} This setting provides a particularly well-motivated realisation of the aforementioned novel UV freeze-in mechanism. In particular, it provides a kination regime together with specific preheating mechanisms able to instantaneously generate a subdominant radiation bath while precluding additional energy injections at later times. We remark that infrared DM freeze-in during kination~\cite{Redmond:2017tja, Visinelli:2017qga}, for which $n<0$, is not affected by this mechanism.

We remark that a background that is stiffer than radiation could alternatively be generated by a coherently oscillating inflaton with a suitable non-trivial potential, e.g. a $\phi^n$ potential with $n>4$ corresponds to $w  = (n-2)/(n+2) > 1/3$~\cite{Turner:1983he}. However, preheating effects due to the self-interaction of the inflaton will result in a transition into a radiation dominated regime within $\mathcal{O}(1)$ $e$-folds~\cite{Lozanov:2016hid, Lozanov:2017hjm}. Thus, a viable $w > 1/3$ background is not expected to be realized in this case.

This paper is structured as follows. In Section~\ref{sec:model} we outline the general quintessential inflation paradigm, paying special attention to the peculiarities of its heating stage. In Section~\ref{sec:DM}, we discuss how the DM relic abundance is produced in the early Universe via the UV freeze-in mechanism. Finally, we provide a summary and some concluding remarks in Section~\ref{sec:end}. Along the paper, we use natural units $\hbar = c = 1$ and a metric signature $(-,\,+,\,+,\,+)$.


\section{Quintessential Inflation}
\label{sec:model}

\noindent Quintessential inflation scenarios employ a single degree of freedom--- dubbed \textit{cosmon}~\cite{Peccei:1987mm} --- to support the early- and late-time accelerated expansion of the Universe \cite{Peebles:1998qn,Spokoiny:1993kt,Brax:2005uf,Hossain:2014xha,Agarwal:2017wxo,Geng:2017mic,Dimopoulos:2017zvq,Rubio:2017gty}. Although non-canonical representations are possible \cite{Wetterich:1987fm, Wetterich:1994bg,  Wetterich:2014gaa, Rubio:2017gty}, and in some cases advisable, the simplest realizations of the paradigm make use of a canonical scalar field $\phi$ with action
\begin{equation}\label{cosmonL}
    S=\int d^4x \sqrt{-g}\left[\frac{\mpl^2}{2}R -\frac{1}{2}\partial_\mu \phi \partial^\mu \phi -U(\phi)\right]\,,
\end{equation}
where $\mpl=(8\pi\,G)^{-1/2}\simeq 2.48\times 10^{18}$ GeV is the reduced Planck mass and $R$ the Ricci scalar. With this standard kinetic sector, a given model within the paradigm is specified by a particular choice of the potential $U(\phi)$, which is assumed to be of the \textit{non-oscillatory} or \textit{runaway} type. Although different potentials (e.g. polynomial \cite{Peebles:1998qn}, exponential \cite{Spokoiny:1993kt,Brax:2005uf,Hossain:2014xha,Geng:2017mic,Rubio:2017gty}, hyperbolic \cite{Agarwal:2017wxo}, etc) give rise to slightly different predictions for the inflationary and dark energy observables, the overall evolution of the Universe in this setting is fairly universal: 

\begin{enumerate}
    \item Inflation takes place at large field values. During this period, the global equation--of--state parameter is close to that of a de Sitter expansion, $w\approx  -1$. As usual, the accelerated expansion of the Universe ends when the slow-roll conditions are violated. 
    \item  Soon after the end of inflation, the system inevitably enters a \textit{kination} or \textit{deflation} regime in which the global equation--of--state parameter approaches the stiff value $w\approx 1$.  
    \item The rapid decrease of the background energy density during kination ($\rho_\phi\sim a^{-6}$) translates into the eventual onset of radiation domination even in those cases in which the depletion of the cosmon field via particle production is still far from complete.
    \item During the first stages of radiation domination, the cosmon condensate \textit{freezes} to a constant value, allowing for the eventual resurgence of its potential energy density as the Universe expands. Provided the production of a DM component, the subsequent evolution of the Universe proceeds according to the usual hot big bang picture. 
    \item When the decreasing energy density of the matter sector becomes comparable with the cosmon energy density, the system settles down to a scaling solution in which the energy density of the cosmon field \textit{tracks} the dominant energy component, see e.g. Ref.~\cite{Rubio:2017gty,Amendola:2018ltt}. 
    \item An exit mechanism from the tracking regime is usually needed to obtain the present dark-energy dominated era.  Although this can be easily implemented in several beyond the SM extensions, a rather natural setup arises if the neutrino--to--electron mass ratio increases with the cosmon field at low redshifts~\cite{Amendola:2007yx, Wetterich:2007kr}. 
    In particular, if the neutrino mass $m_\nu$ depends on $\phi$ for fixed electron mass, it induces an additional term in the  Klein-Gordon equation for the scalar field~\cite{Wetterich:1994bg,Fardon:2003eh,Brookfield:2005bz} that becomes active \textit{only} when neutrinos become non-relativistic, namely
    \begin{equation}
        \ddot \phi +3 H\dot \phi+ U_{,\phi}=\frac{\beta}{M_P}\left(\rho_{\nu}-3 p_{\nu}\right)\,,
    \end{equation}
    with 
    \begin{equation}
        \beta\equiv -M_P\frac{\partial}{\partial\phi}\ln m_{\nu}(\phi)\,,
    \end{equation} 
    and $\rho_\nu$ and $p_\nu$ the neutrino energy density and pressure. For negative $\beta$ (growing neutrino masses), the evolution of the cosmon field stops at the approximate redshift of the relativistic-to-non-relativistic neutrino transition ($z\approx 5$)~\cite{Mota:2008nj}, terminating the tracking regime and leading to a background cosmology rather close to $\Lambda$CDM afterwards.
\end{enumerate}

\subsection{Heating}
\label{sec:reheat}

Many heating mechanisms for quintessential inflation have been proposed in the literature~\cite{Ford:1986sy, Damour:1995pd, Peebles:1998qn, Felder:1999pv, Feng:2002nb, BuenoSanchez:2007jxm, Rubio:2017gty, Nakama:2018gll, Dimopoulos:2018wfg}. The simplest, albeit highly inefficient, possibility \cite{Rubio:2017gty, Figueroa:2018twl} is the inevitable production of scalar particles in an expanding Universe \cite{Spokoiny:1993kt,Ford:1986sy,Damour:1995pd}. This mechanism applies, for instance, to the SM Higgs or to other (non-conformally coupled) scalar fields appearing in grand unified theories, but not to gauge bosons and chiral fermions given their Weyl-invariant character. Particle production can alternatively take place via direct couplings of the cosmon field to matter \cite{Felder:1999pv,Rubio:2017gty} or non-minimal couplings to gravity~\cite{Dimopoulos:2018wfg,Opferkuch:2019zbd,Bettoni:2019dcw}. The details of the heating process are, however, not important from an observational perspective. To see this explicitly let us assume that any of the above mechanisms is able to produce a partial depletion of the cosmon condensate by the creation of relativistic particles during a short period of time at the very onset of kination. The efficiency of the process can be parametrized by a \textit{heating efficiency} parameter~\cite{Rubio:2017gty}
\begin{equation}\label{thetadef}
    \Theta\equiv \frac{\rho_{\rm R}^{\rm kin}}{\rho^{\rm kin}_\phi}\,,
\end{equation}
with $\rho^{\rm kin}_\phi$ and $\rho_{\rm R}^{\rm kin}$ the energy densities of the cosmon field and its decay products at that time. Using the scaling of the different energy components during the kination epoch ($\rho_\phi\sim a^{-4}$ and $\rho_R\sim a^{-6}$) together with entropy conservation, we can easily relate the efficiency parameter \eqref{thetadef} to the \textit{reheating temperature} at which the energy density of the created particles equals that of the cosmon field ($\rho_{\rm R}^{\rm RH}=\rho_\phi^{\rm RH}$), namely
\begin{equation}\label{eq:Trh}
    \Trh\equiv \left(\frac{30}{\pi^2 \gsrh}\,\rho_{\rm R}^{\rm RH}\right)^{1/4}\,,
\end{equation}
with  $\gsrh$ the effective number of relativistic degrees of freedom at that time. After some trivial algebra, we get 
\begin{equation} \label{Treh0}
    T_{\rm RH}
    =\left(\frac{\gsskin}{\gssrh}\right)^{1/3}\Theta^{1/2} \,\Tmax\,,
\end{equation}
with $\gsskin$ and $\gssrh$ denoting entropic degrees of freedom and 
\begin{equation}\label{eq:Tmax}
    \Tmax\equiv \left(\frac{30}{\pi^2\,\gskin}\,\rho_{\rm R}^{\rm kin}\right)^{1/4}
\end{equation}
the maximum temperature of the created particles at the onset of kination under the assumption of instantaneous thermalization. We observe then that the smaller the heating efficiency, the longer the kination epoch and the larger the difference between the maximal radiation temperature and the proper reheating temperature. In particular, accounting for all SM degrees of freedom at temperatures higher than the top-quark mass ($\gskin=\gsskin=\gssrh=106.75$), we have 
\begin{eqnarray}
    \Tmax
        &\simeq& 2.5 \times 10^{12}~\GeV\left(\frac{\Theta}{10^{-8}}\right)^\frac14\left(\frac{\Hkin}{10^{11}~\GeV}\right)^\frac12,\label{Tmax}\\
    \Trh 
        &\simeq& 2.5\times 10^8~\GeV\left(\frac{\Theta}{10^{-8}}\right)^\frac34\left(\frac{\Hkin}{10^{11}~\GeV}\right)^\frac12.\label{Treh}
\end{eqnarray}
The upper limit on the scale of inflation $H_{*} < 2.5\times10^{-5}\, \mpl$~\cite{Akrami:2018odb} implies thus a bound $\Tmax \leq 6 \times 10^{11}~\GeV \times \Theta^{1/4}$ on the maximal temperature.

All the heating details are effectively encoded in the value of $\Theta$. For a fiducial Hubble rate $H_{\rm kin }\sim 10^{11}\,{\rm GeV}$, the heating efficiency per degree of freedom varies between $\Theta\sim 10^{-19}$ in gravitational heating scenarios \cite{Ford:1986sy,Damour:1995pd,Peebles:1998qn} and $\Theta\sim {\cal O}(1)$ in heating scenarios involving direct couplings of the cosmon field to matter \cite{Felder:1999pv,Rubio:2017gty}. The main phenomenological restriction on its value comes from the requirement that the primordial spectrum of gravitational waves generated during inflation does not become excessively enhanced during kination as to spoil the relative abundance of light elements. Taking into account the integrated  big bang nucleosynthesis (BBN) constraint on the gravitational wave density fraction $\Omega_{\rm GW}$ \cite{Maggiore:1999vm,Caprini:2018mtu},
\begin{equation}\label{GWbound1}
h^2\int_{k_{\rm BBN}}^{k_{\rm kin}}\Omega_{\rm GW}(k)\, d\ln k \lesssim 1.12\times 10^{-6}\,, 
\end{equation}
this translates into a lower bound \cite{Rubio:2017gty}
\be\label{GWbound2}
    \Theta\gtrsim          10^{-16}\left(\frac{H_{\rm kin}}{10^{11}\,{\rm GeV}}\right)^2\,.
\ee
This restriction cannot be generically satisfied in gravitational heating scenarios without invoking  ${\cal O}(100)$ scalar fields  beyond the SM content. On top of that, even if these large number could be accommodated in a given SM extension, gravitational particle production should not be considered a completely satisfactory heating mechanism since the inclusion of sizable number of light fields in the spectrum could lead to the production of large isocurvature perturbations or secondary inflation periods \cite{Felder:1999pv}. These problems are, however, absent in the presence of \textit{direct} couplings between the cosmon field and matter considered in the next section.

\subsection{A Worked-out Example}
\label{sec:example}

To illustrate the general considerations above, let us consider the following interaction between the cosmon field $\phi$ and a scalar matter field $h$,
\begin{equation}\label{HLag0}
S_I=-\int d^4x \sqrt{-g} \left[\frac{1}{2}(\partial h)^2 + \frac{1}{2} m_h^2(\phi)\,h^2\right].
\end{equation}
In order for this scenario to be phenomenologically viable, the function $m_h^2(\phi)$ in this expression should be $i)$ large enough during inflation as to retain the single field inflationary dynamics, $ii)$ vary  rapidly at the end of inflation to heat the Universe  via adiabaticity violations, and $iii)$ decrease monotonically with time in order to avoid strong backreaction effects at large $\phi$ values.  A simple choice satisfying all these criteria is%
\footnote{Note that this a priori unconventional behaviour is expected in models  where quintessential inflation is associated with the emergence of quantum scale symmetry in the vicinity of UV and IR fixed points~\cite{Wetterich:2014gaa,Rubio:2017gty}. In this type of variable gravity scenarios \cite{Wetterich:2013jsa}, rapid variations of $m^2_h(\phi)$ are only expected to occur in a crossover regime where the dimensionless couplings and mass ratios of matter fields evolve from their UV to IR values. If the field $h$ in Eq.~\eqref{mass} is identified with the SM Higgs, the decoupling from $\phi$ at late times encodes the approach to the SM IR fixed point, as required by the constraints on the variation of the Fermi to Planck mass ratio since nucleosynthesis \cite{Uzan:2010pm,Wetterich:2003qb}.}
\begin{equation}\label{mass}
m_h^2(\phi)=\begin{cases} 
      g^2 \phi^2 & \quad\text{ for }\phi\leq 0\,, \\
      \tilde m_h & \quad\text{ for }\phi >  0\,,
   \end{cases}
\end{equation}
with $\tilde m_{h}$ being a constant.  Alternative choices sharing the features described in $i)$, $ii)$ and $iii)$ could be used without significantly modifying the conclusions below. For instance, one could introduce a  parameter $\phi_l$ encoding the \textit{timing} of the  transition by simply replacing $\phi$ by $\phi-\phi_l$ or consider smoothing the transition at $\phi=0$ by using some interpolation function, see e.g. Ref.~\cite{Rubio:2017gty}. For the sake of simplicity, we will restrict ourselves to the simplest form \eqref{mass} in the following estimates. 

To understand how the effective coupling in Eq.~\eqref{HLag0} gives rise to particle production, let us consider the $h$-field equation of motion in Fourier space for an homogeneous field configuration $\phi=\phi(t)$, namely 
\be\label{motion}
\ddot h_k+3 H \dot h_k+\left(\frac{k^2}{a^2}+m_h^2(\phi)\right)h_k=0\,,
\ee
with $k$ the momentum.  After eliminating the Hubble friction in this expression term by performing a field redefinition  $h_k\rightarrow a^{-3/2} h_k$, we obtain a time-dependent harmonic oscillator equation
\begin{equation}\label{eq:modes} 
\ddot h_k+\omega_k^2(t)\, h_k=0\,,\hspace{20mm} \omega_k^2(t)=\frac{k^2}{a(t)^2}+m^2_h(t)\,,
\end{equation}
where we have omitted a subdominant time-dependent mass contribution $\Delta_a\equiv -\frac{3}{4}\frac{\dot a^2}{a^2}-\frac{3}{2}\frac{\ddot a}{a}$ responsible from the usual gravitational particle production in an expanding Universe. The solution of this differential equation can be  well-described by a Wentzel–Kramers–Brillouin approximation whenever the adiabaticity condition $\dot \omega_k/\omega_k^2 \ll 1$ is satisfied. When violated, particle production takes place in the infrared part of the spectrum~\cite{Lozanov:2019jxc}. 

At small $k$ values, the violation of the adiabaticy condition can be safely approximated by $|\dot m_{h}| \gtrsim m^2_{h}$ or equivalently by $g^2 \phi^2 \lesssim g|\dot\phi_0|$, with $\vert \dot \phi_0\vert $ the cosmon velocity at zero crossing. Solving this expression for $\phi$, we observe that particle production takes place in a very narrow interval $\Delta \phi \sim (|\dot\phi_0|/ g)^{1/2}$ around $\phi=0$, being the production essentially instantaneous for sufficiently large couplings, $\Delta t \sim  \phi_*/|\dot\phi_0| \sim (g |\dot\phi_0|)^{-1/2}$. The typical momentum of the created particles follows directly from the uncertainty principle, $\Delta k \sim (\Delta t)^{-1} \sim  (g |\dot\phi_0|)^{1/2}$ and coincides with the one obtained by properly solving the mode equation \eqref{eq:modes} in the WKB approximation. Indeed, as shown explicitly in the seminal paper \cite{Kofman:1997yn}, the occupation number of $h$ particles after a single zero crossing is given by
\begin{equation}\label{number}
    n^{\rm kin}_k 
    = \exp\left(- \frac{ \pi k^2}{g|\dot\phi_0|}\right)\,.   
\end{equation}
Assuming the decay products to be ultra-relativistic ($g|\dot\phi_0|\gg \tilde m_h$), this corresponds to an \textit{instantaneous} radiation energy density generation  \begin{equation}\label{eq:energyh}
\rho^{\rm kin}_R=\int\frac{d^3k}{(2\pi)^3}\, \omega_k\,  n_k\simeq \frac{g^2}{4 \pi ^4} \vert \dot \phi_0\vert^2 \,.
\end{equation} 
\textit{No additional energy injections take place afterwards} since the adiabaticity condition is never else violated. Combining Eq.~\eqref{eq:energyh} with its cosmon counterpart,
\begin{equation}
    \rho_{\phi}^{\rm kin}=3\mpl^2\,H_{\rm kin}^2\,,
\end{equation}
and assuming the thermalization process in the radiation sector to take place instantaneously, we obtain the following estimate for the heating efficiency \eqref{thetadef} in this particular example, 
\begin{equation}
\Theta\simeq 2 \times 10^{-8} \left(\frac{g}{0.02}\right)^2 \left(\frac{10^{11} \, {\rm GeV}}{H_{\rm kin}}\right)^2\left(\frac{\vert \dot \phi_0\vert }{10^{-8} \mpl^2}\right)^2   \,. 
\end{equation} 
The heating process depends therefore on the scale of inflation, the velocity of the field at zero crossing and the strength of the cosmon coupling to matter, being the results completely independent of the particle spin. This allows to extend the above estimates to fermionic species. Note that this would not be the case in oscillatory scenarios since the adiabaticity condition is violated there periodically, leading to bosonic enhancement effects~\cite{Kofman:1997yn}. 
For completeness, let us finally mention that if the adiabaticity condition is violated in $n_a$ channels,  the above quantities are rescaled as $\Theta\rightarrow n_a\, \Theta$, $T_{\rm max }\rightarrow \sqrt{n_a}\,  T_{\rm max}$ and  $T_{\rm RH}\rightarrow \sqrt{n_a}\,  T_{\rm RH}$, respectively.

\section{Dark Matter Production} 
\label{sec:DM}

The evolution of the DM number density $n$ is given by the Boltzmann equation
\begin{equation}\label{eq:DMBE0}
	\dot n + 3H\,n  = -\sv\left(n^2-n_\text{eq}^2\right),
\end{equation}
with $n_\text{eq} = \frac{g\,\zeta(3)}{\pi^2}\,T^3$ the equilibrium DM number density for relativistic particles, $\zeta(3)\simeq 1.2$  the Riemann zeta function of 3 and $g$ the number of DM degrees of freedom.%
\footnote{An additional multiplicative factor of $3/4$ is present for fermionic DM. Hereafter bosonic DM will be assumed.
Furthermore, thermalization and number-changing processes within the dark sector can have a
strong impact on the DM relic abundance. In particular, they can enhance the DM abundance by several
orders of magnitude\cite{Bernal:2020gzm}.}
As preempted in Eq.~\eqref{eq:sv}, we will assume the thermally averaged DM annihilation cross section to be a function of the thermal bath temperature $T$, namely
\be\label{eq:sv2}
	\sv=\frac{T^n}{\Lambda^{n+2}}\,.
\ee
This cross section could be generated by non-renormalizable operators of mass dimension $5+n/2$, with $n$ even and non-negative. The scale $\Lambda$ stands for the cutoff of the effective field theory and could be interpreted as a proxy of the mediator mass connecting the dark and visible sectors. In order to ensure the validity of this effective operator description, we will require $\Lambda$ to be the highest scale in our computations. On top of that, we will concentrate on scenarios displaying a scale hierarchy $m\ll T\ll\Lambda$, with $m$ the DM mass.

To track the evolution of the DM density in those cases in which the SM entropy is conserved,\footnote{We remark that it is a priori possible to construct scenarios where entropy is injected into the SM thermal bath throughout kination. On general grounds, this tends to slow down the temperature scaling as compared to Eq.~\eqref{eq:sT}. For example, a constant conversion rate of the inflaton energy density into radiation would generate a scaling law $\rho_{R} \propto \sqrt{\rho_{\phi}} \propto a^{-3}$ prior to radiation domination~\cite{Bernal:2019mhf}. We do not consider this possibility here.} it is convenient to rewrite  Eq.~\eqref{eq:DMBE0} in terms of the so-called DM yield $Y\equiv n/s$, with 
\begin{equation}\label{eq:sT}
	s(T)\equiv \frac{2\pi^2}{45}\,\gss(T)\,T^3\,,
\end{equation}
the SM entropy density and $\gss(T)$ the effective number of relativistic degrees of freedom contributing to it~\cite{Drees:2015exa}. Taking into account the fact that in freeze-in production DM never reaches its equilibrium distribution, i.e. $Y\ll Y_\text{eq}$ , Eq.~\eqref{eq:DMBE0} becomes
\begin{equation}\label{eq:DMBE}
	\frac{dY}{dT} =\frac{\sv\,s}{H\,T}\left(Y^2-Y_\text{eq}^2\right)\simeq-\frac{\sv}{s\,H\,T}\,n_\text{eq}^2\,.
\end{equation}

In the sudden-decay approximation for the inflaton field, the scale $\Trh$ coincides with the maximal temperature reached by the SM thermal bath. On top of that, the Universe is customarily assumed to be dominated by the SM radiation prior to matter-radiation equality.
In this standard cosmological scenario, the yield equation \eqref{eq:DMBE} can be analytically solved. In particular, its asymptotic solution at $T\ll\Trh$ can be written as
\begin{equation}\label{eq:Yrh_standard}
	Y_0\simeq \frac{135\,\zeta(3)^2}{2\pi^7(n+1)}\sqrt{\frac{10}{\gs}}\frac{g^2}{\gss}\frac{\mpl}{\Lambda^{n+2}}\Trh^{n+1},
\end{equation}
up to small corrections associated to the temperature dependence of $\gs$ and $\gss$. Within this approximation, the bulk of the DM abundance is produced at $T\simeq\Trh$, a characteristic of the UV freeze-in.

The situation is completely different in those scenarios involving an expansion era with a stiff equation-of-state parameter $w>1/3$ and no continuous entropy injection into the SM constituents, such that the photon temperature maintains the standard scaling $T\propto a^{-1}$.  Under these conditions, and taking into account the relation $H\propto a^{-\frac32(1+w)}$, the yield equation~\eqref{eq:DMBE} can be analytically integrated up to small $\gs$ and $\gss$ variations to obtain\footnote{For $n=n_c$ the bracket in Eq.~\eqref{eq:GeneralYrh} becomes $\ln(\Tmax/\Trh)$.}  
\begin{equation}\label{eq:GeneralYrh}
	Y^{\rm stiff}_0
	\simeq \frac{135\,\zeta(3)^2}{2\pi^7}\sqrt{\frac{5}{\gs}}\frac{g^2}{\gss}\frac{\mpl\,\Trh^{n_c+1}}{\Lambda^{n+2}}\left[\frac{\Tmax^{n-n_c}-\Trh^{n-n_c}}{n-n_c}\right]\,,
\end{equation}
where the critical threshold is given by
\be
    n_c\equiv\frac32(w-1)\,.
\ee
Regarding cross sections~\eqref{eq:sv2} with $n>n_c$, the freeze-in production is highly enhanced with respect to the standard case \eqref{eq:Yrh_standard}.
This parametric enhancement can be described by a boost factor~\cite{Garcia:2017tuj, Bernal:2019mhf}
\begin{equation}
    B\equiv \frac{	Y^{\rm stiff}_0}{Y_0} \simeq
    \frac{1}{\sqrt{2}}\frac{n+1}{n-n_c}\left(\frac{\Tmax}{\Trh}\right)^{n-n_c}
    \propto \Theta^{-\frac{n-n_c}{2}},
\end{equation}
for $n>n_c$ and $\Tmax \gg \Trh$. 

\begin{figure}[t!]
	\begin{center}
		\includegraphics[height=0.33\textwidth]{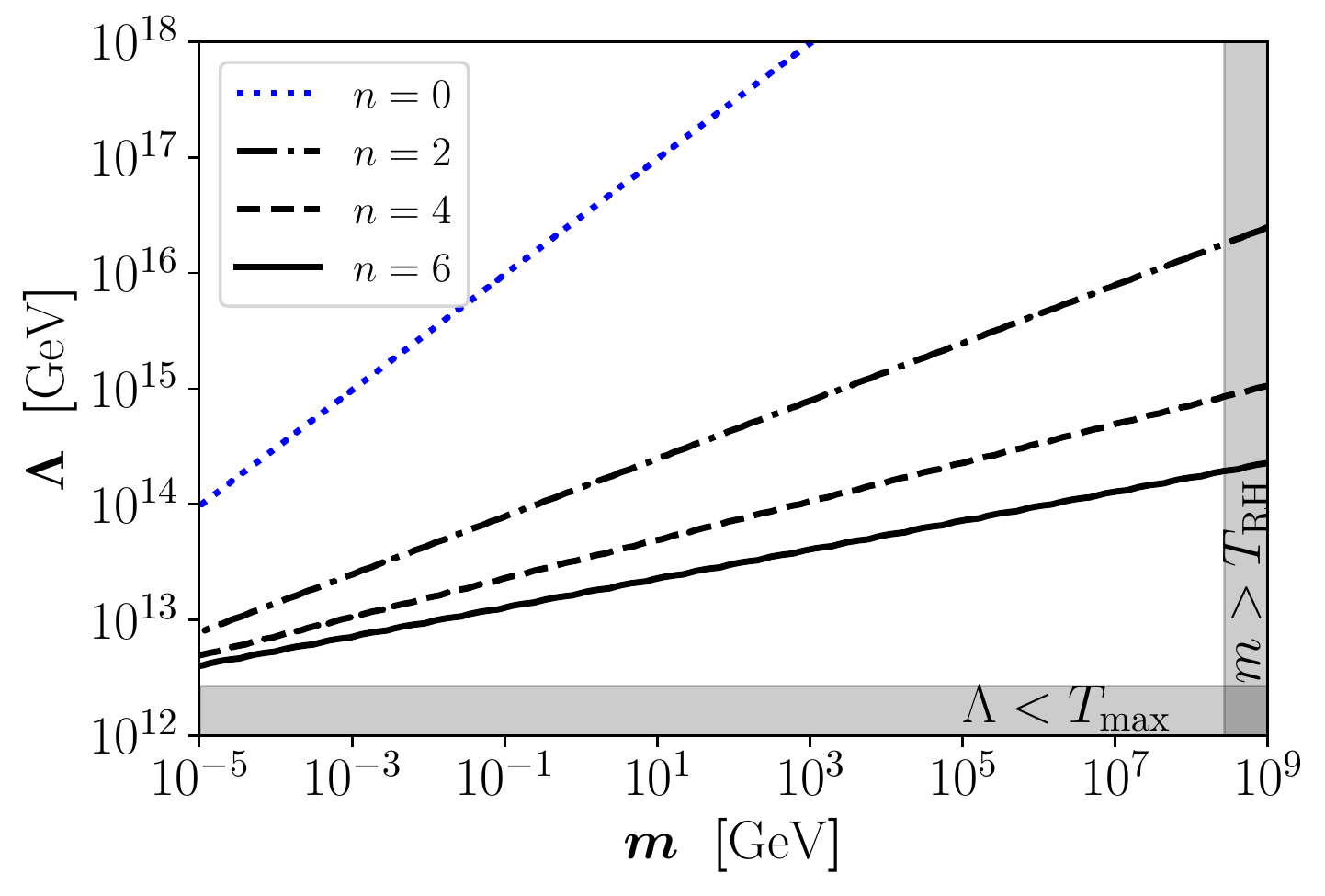}
		\includegraphics[height=0.33\textwidth]{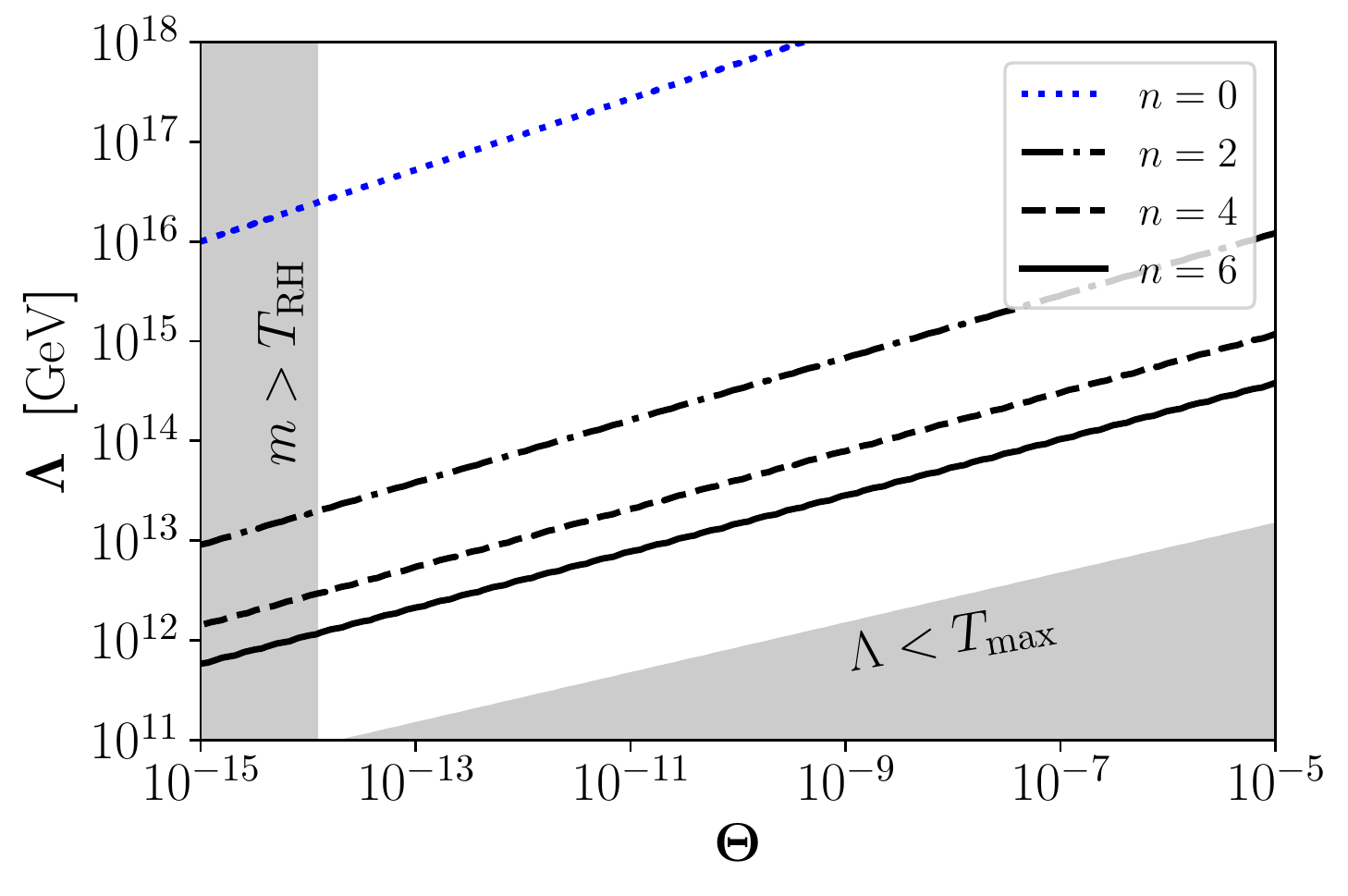}
		\caption{Parameter space reproducing the observed DM relic abundance for different values of $n$, $\Hkin=10^{11}$~GeV and $\Theta=10^{-8}$ (left panel) or $m=10^4$~GeV (right panel).
        The gray regions correspond to $\Lambda<\Tmax$ and $m>\Trh$.
		}
		\label{fig:DM}
	\end{center}
\end{figure}

The above discussion applies to a broad class of cosmological scenarios where, after the heating period, the Universe is dominated by a component that is stiffer than radiation, i.e. with an equation-of-state parameter $w>1/3$. In particular, NO models provide just one realization of this general framework. During the kination period at $\Tmax>T>\Trh$, the effective equation-of-state parameter is given by $w\simeq 1$ and therefore $n_c=0$. In this limit and for $n>0$, the yield in \eqref{eq:GeneralYrh} becomes%
\footnote{An interesting example is the $s$-channel exchange of a graviton in the annihilation of SM states. In this case $n=2$. This process gives an irreducible contribution to the total DM relic abundance~\cite{Garny:2015sjg, Tang:2017hvq, Garny:2017kha, Bernal:2018qlk}.
The corresponding interaction rate $\sv=\alpha\,T^2/\mpl^4$, with $\alpha$ being an $\mathcal{O}(10^{-1})$ parameter, yields
\be
	Y_0 
	\simeq \frac{135\,\zeta(3)^2\,\alpha}{4\pi^7}\sqrt{\frac{5}{\gs}}\frac{g^2}{\gss}\frac{\Trh\,\Tmax^2}{\mpl^3}\,
	\lesssim \alpha\,g^2 \, 5 \times 10^{-25}.
\ee
This contribution is, however, negligible due to the constraint $\Tmax \leq 2.5 \times 10^{-7} \mpl$ from the scale of inflation.}
\begin{equation}\label{eq:Yrh}
	Y^{w\simeq 1}_0 \simeq \frac{135\,\zeta(3)^2}{2\pi^7}\sqrt{\frac{5}{\gs}}\frac{g^2}{\gss}\frac{\mpl\,\Trh}{\Lambda^{n+2}}\left[\frac{\Tmax^n-\Trh^n}{n}\right]\,,
\end{equation}
with $\Tmax$ and $\Trh$ completely determined by the scale of inflation  $\Hkin$ and the heating efficiency $\Theta$, cf.~Eqs.~\eqref{Tmax} and~\eqref{Treh}. An example of the parameter space reproducing the observed DM relic abundance for different values of $n$, $\Hkin=10^{11}$~GeV and $\Theta=10^{-8}$ (left panel) or $m=10^4$~GeV (right panel) is shown in Fig.~\ref{fig:DM}. The gray areas for $\Lambda<\Tmax$ and $m>\Trh$ correspond to cases where our effective field approach fails.
In order to compute $\Lambda$, the DM yield has been held fixed so that $m\,Y_0 = \Omega_\text{DM} h^2 \, \frac{1}{s_0}\,\frac{\rho_c}{h^2} \simeq 4.3 \times 10^{-10}$~GeV, with $\rho_c \simeq 1.1 \times 10^{-5} \, h^2$~GeV/cm$^3$  the critical energy density, $s_0\simeq 2.9\times 10^3$~cm$^{-3}$ the entropy density at present and $\Omega_\text{DM} h^2\simeq 0.12$~\cite{Aghanim:2018eyx}.
Hence, for a DM with mass $m$ and $g$ internal degrees of freedom, it is required that
\be
	\Lambda \simeq \Trh \left[  4\times 10^{23}  g^2 \,\frac{m}{\Trh}\, \frac{(\Tmax/\Trh)^{n}-1}{n}\right]^\frac{1}{n+2}.
\ee
For the benchmark chosen in Fig.~\ref{fig:DM}, the observed DM relic abundance can be obtained within a wide range of DM masses, spanning from the keV to the PeV scale, and for scales of new physics well bellow the Planck mass scale.

A common feature of any specific realization of our set-up is that the SM-DM interactions arise through higher dimensional effective operators at scales above the maximal bath temperature \eqref{eq:Tmax}, implying that $\Lambda > 10^{11}~\GeV$. As an exception, with an exact kination epoch, a boost factor may be realized even with renormalizable interactions, because the critical exponent vanishes, $n_c = 0$. The required constant cross section can be realized, for example, via interacting gauge bosons~\cite{Gabrielli:2015hua}.

The DM-SM interactions predicted by our set-up are currently out of experimental reach due to their weakness, as usual for UV freeze-in scenarios. Moreover, although a detection of only feeble interactions between the dark and the visible sector tends to favour freeze-in scenarios, this alone would not provide a model independent confirmation of all our assumptions, which include a kination era and a heating efficiency depending on the specifics of the underlying particle physics model. Additional observables, sensitive to the expansion in the early Universe, e.g. gravitational waves~\cite{Bernal:2019lpc, Figueroa:2019paj}, or to the maximal temperature of the heat bath, would be most probably required to confirm the proposed scenario. 

\section{Conclusions} 
\label{sec:end}

We considered UV freeze-in in cosmologies where the inflationary stage is followed by a expansion era dominated by a fluid component stiffer than radiation, that is, with an equation-of-state parameter $w>1/3$. If a fraction of this fluid energy density is transferred to the SM particles in one way or another, the resulting SM bath will eventually dominate the energy budget of the Universe due to the rapid dilution of the stiff energy component. As a concrete realization of this general framework, we considered quintessential NO inflation scenarios able to provide the initial SM thermal bath through a sudden event of preheating at the onset of kination.

In the scenario considered in this paper, the DM abundance can be significantly enhanced as compared to the sudden decay approximation, where the abundance is determined by the reheating temperature setting the onset of radiation domination. The reason for this enhancement is two-fold. First, during the period where the thermal bath is subdominant, the maximal temperature can greatly exceed the reheating temperature. Thus, the bulk of DM will be produced before reheating. We note that this is a common feature of UV freeze-in models, typically associated with the non-instant (perturbative) decay of the inflation field (see e.g. Ref.~\cite{Bernal:2019mhf}). Second, an additional boost is provided by the fact that the inflaton energy density becomes subdominant just because it is diluted faster than radiation, \textit{not} because it decays. This avoids additional entropy injections into the SM thermal bath, preventing the dilution of DM abundance which would otherwise occur in models in which the inflaton is unstable and completely depleted by a continuous perturbative decay. The second feature is a novel characteristic of the UV freeze-in scenarios considered here. In particular, during kination, we find a strong enhancement if DM production happens via cross sections scaling as $\sv\propto T^n$ with $n\geq 0$.

The above results are based on two ingredients regarding the production of the SM plasma and its subsequent thermalization: 
\begin{enumerate}[i)]
\item \textit{Non-linear effects generically appearing in oscillatory models are totally absent in the NO quintessential inflation scenarios considered here}. These effects include, among others, a parametric-resonance stage leading to the exponential growth of fluctuations~\cite{Kofman:1997yn,Greene:1997fu,Berges:2002cz}, a combined preheating delay \cite{GarciaBellido:2008ab,Rubio:2015zia,Repond:2016sol}, a nonlinear fragmentation of the inflation field \cite{Felder:2006cc}, and a slow approach to equilibrium via driven turbulence and self-similar evolution \cite{Micha:2002ey,Micha:2004bv}. We emphasize that this is a generic sequence of events for non-quadratic inflationary potentials, even in the absence of couplings to other fields. The associated temporal scales vary between many $e$-folds in pure power-law inflationary models and less than an $e$-fold in models where the potential flattens out at distances $\Delta\phi<M_P$ from its minimum \cite{Repond:2016sol,Lozanov:2016hid,Lozanov:2017hjm,Hasegawa:2017iay,Krajewski:2018moi,Lozanov:2019ylm,Rubio:2019ypq}. As a result, it is difficult to realize a sustained background evolution that is diluted faster than radiation by using a coherently oscillating inflaton field---associated for instance to a $\phi^n$ potential with $n>4$~\cite{Turner:1983he}---since its self-interactions will generically trigger a transition to radiation domination via inflaton fragmentation~\cite{Lozanov:2016hid, Lozanov:2017hjm}. Furthermore, the population of the SM sector in oscillatory models (see e.g. Ref.~\cite{Kaneta:2019zgw}) is not generically realized through the perturbative decay of the inflaton field, but rather through a complicated depletion process most likely requiring the use of lattice simulations or realistic fitting formulas directly extracted from them.

\item \textit{The thermalization process leading to the SM plasma is taken to be instantaneous.} This is expected to be a good approximation if the $h$ field in Section \ref{sec:example} is identified with the SM Higgs field, but non-thermal effects could a priori play a role in more involved scenarios, increasing in some cases the yield by several orders of magnitude~\cite{Garcia:2018wtq}. This would change the results presented here quantitatively, but non qualitatively.
\end{enumerate}
Given these ingredients, the DM production mechanism presented in this paper is rather general and not restricted to any particular realization of the quintessential inflationary scenario or to the specifics of the heating process. In particular, it applies to any cosmological setting involving a post-inflationary era with equation-of-state parameter $1/3 < w\leq 1$, irrespective of its origin. On top of that, our framework can be easily extended to accommodate additional phenomenology such as $i)$ the generation of the observed baryon asymmetry  \cite{Bettoni:2018utf}, $ii)$ the production of  short-lived topological defects able to generate a potentially detectable gravitational wave background~\cite{Bettoni:2018pbl, Bettoni:2019dcw}, and $iii)$ the confinement of the created DM particles into black holes~\cite{Amendola:2017xhl} and primordial DM halos~\cite{Savastano:2019zpr}.

\section*{Acknowledgments}
The authors thank Mustafa Amin, Dario Bettoni and Tommi Markkanen for useful discussions and comments on the manuscript. HV is supported by the Estonian Research Council grants MOBTTP135, PRG803 and MOBTT5 and by the EU through the European Regional Development Fund CoE program TK133 ``The Dark Side of the Universe".
Additionally, this project has received funding from the European Union's Horizon 2020 research and innovation program under the Marie Sk\l{}odowska-Curie grant agreements 674896 and 690575, and from Universidad Antonio Nari\~no grants 2018204, 2019101 and 2019248.
NB is partially supported by Spanish MINECO under Grant FPA2017-84543-P.

\bibliographystyle{JHEP}

\bibliography{biblio}

\end{document}